\documentclass[twocolumn,preprintnumbers,amsmath,amssymb]{revtex4}

\usepackage{graphicx}
\usepackage{dcolumn}
\usepackage{bm}

\begin{document}







\def\beq{\begin{equation}}
\def\eeq{\end{equation}}
\def\bea{\begin{eqnarray}}
\def\eea{\end{eqnarray}}
\def\ben{\begin{enumerate}}
\def\een{\end{enumerate}}
\def\la{\langle}
\def\ra{\rangle}
\def\a{\alpha}
\def\b{\beta}
\def\g{\gamma}\def\G{\Gamma}
\def\d{\delta}
\def\e{\epsilon}
\def\phi{\varphi}
\def\k{\kappa}
\def\l{\lambda}
\def\m{\mu}
\def\n{\nu}
\def\o{\omega}
\def\p{\pi}
\def\r{\rho}
\def\s{\sigma}
\def\t{\tau}
\def\L{{\cal L}}
\def\S{\Sigma }
\def\gsim{\; \raisebox{-.8ex}{$\stackrel{\textstyle >}{\sim}$}\;}
\def\lsim{\; \raisebox{-.8ex}{$\stackrel{\textstyle <}{\sim}$}\;}
\def\gtrsim{\gsim}
\def\lessim{\lsim}
\def\loc{{\rm local}}
\def\vm{v_{\rm max}}
\def\bh{\bar{h}}
\def\del{\partial}
\def\nab{\nabla}
\def\half{{\textstyle{\frac{1}{2}}}}
\def\fourth{{\textstyle{\frac{1}{4}}}}

\title{Static post-Newtonian equivalence of GR and\\ 
gravity with a dynamical preferred frame}

\author{Christopher Eling}
 \email{cteling@physics.umd.edu}
 \affiliation{Department of Physics, University of Maryland\\ 
College Park, MD 20742-4111 USA}
\author{Ted Jacobson}%
 \email{jacobson@physics.umd.edu}
 \altaffiliation[Present address: ]
{Insitut d'Astrophysique de Paris, 98 bis Bvd.~Arago, 75014 Paris, FRANCE}
\affiliation{Department of Physics, University of Maryland\\ 
College Park, MD 20742-4111 USA}


\begin{abstract}

A generally covariant extension of general relativity (GR)
in which a dynamical unit timelike vector field 
is coupled to the metric is studied
in the asymptotic weak field limit of
spherically symmetric static solutions. 
The two post-Newtonian parameters known as the Eddington-Robertson-Schiff
parameters are found to be identical to those in the case of pure GR,
except for some non-generic values of the
coefficients in the Lagrangian.

\end{abstract}
\maketitle

\section{Introduction}
\label{Intro}

Over the past several years quantum gravity considerations have led
a number of researchers to contemplate violations of Lorentz
invariance.  Such violations are usually studied within the
realm of particle physics, but they would also have implications
for gravitation. Fixed background tensors breaking Lorentz symmetry
also break the general covariance of general relativity (GR),
generically resulting in over-determined equations of motion.
Together with a theoretical bias for preserving general covariance,
this leads one to consider Lorentz breaking introduced by
{\it dynamical}  tensor fields.

The background value of a given dynamical tensor field ``spontaneously"
breaks Lorentz symmetry, either because of a potential that has a Lorentz
violating (LV)  minimum, or because the tensor is somehow constrained not
to vanish. The simplest situations arise from scalar fields whose non-zero
gradient is an LV vector, and vector fields. Vector fields without any potential or constraint were considered in the early 1970's
by Will, Nordvedt, and Hellings in the spirit of alternate
theories of gravity~\cite{willnord,nordwill,hellnord,willbook}. 
Vector fields with a potential
leading to LV were studied by  Kosteleck\'{y} and Samuel~\cite{KosSam}
motivated by string theory considerations, and
by Clayton and Moffat~\cite{moffatv}
motivated by the notion that a dynamically varying speed of
light might solve some cosmological problems (see also
\cite{moffats,Bassett:2000wj}).

A vector field constrained to have a fixed timelike
or spacelike length breaks local Lorentz symmetry
in every configuration, much as a nonlinear sigma
model spontaneously breaks gauge invariance.
In the timelike case the residual symmetry is the 3D
rotation group, while in the spacelike case it is the 2+1
Lorentz group.
A particularly simple example of such a theory was considered
by Kosteleck\'{y} and Samuel~\cite{KosSam} and also studied by
Jacobson and Mattingly~\cite{jacobmatt}.
In this example, the action for the covariant vector field is just the
square of the exterior derivative. Like the Maxwell action
for the vector potential, this is independent of the connection
components.

The most general action with up to two derivatives
of a unit vector field has four terms.
Three of these
terms were included in the original study of
Will and Nordvedt~\cite{willnord},
(which did not include any constraint on the length of the
vector) and all four were written down and studied
using the tetrad formalism by
Gasperini~\cite{Gasperini}, who broke the
local Lorentz symmetry  by including
in the action terms referring to a fixed ``internal" unit timelike
vector. The same theory in the metric formalism
was written down in \cite{jacobmatt},
and the linearized wave solutions in one special case
(corresponding to the case focused on in \cite{willnord})
were reported in \cite{mattjacob}.
The general constrained vector-tensor theory
is quite complicated due to the derivative coupling
to the vector field which includes connection
components. Therefore the action for the vector in fact
modifies the kinetic terms for the metric as well.
Thanks to the unit timelike constraint, the
vector has only three degrees of freedom, all corresponding
to spacelike variations. Thus, unlike in other vector
theories without gauge invariance, 
problems with negative energy modes need not arise.

We are interested in the observational signatures
and constraints on the parameters in the general
constrained vector-tensor theory.
These can be obtained from the parametrized post-Newtonian
(PPN) parameters,
wave phenomena, and strong field effects.
The PPN parameters for the general unconstrained vector-tensor
theory were found by Will~\cite{willbook}. Those results
do not directly apply to the constrained case, hence the
analysis must be carried out anew. In this paper
we begin that process, restricting at first to the static
PPN  parameters, i.e. the Eddington-Robertson-Schiff
parameters $\b$ and $\g$, which are the only ones that
do not vanish when the isolated system is at rest with respect
to the asymptotic rest frame defined by the vector field.
We find that for generic values of
the coefficients in the Lagrangian these parameters take the
same values as in GR. This indicates that to observationally
bound the coefficients in the lagrangian one
must consider higher order PPN contributions, 
preferred frame effects associated with the
motion of the solar system relative to the asymptotic rest frame
of $u^a$, and/or radiation or other effects.

We use metric
signature $({+}{-}{-}{-})$ and units with $c=1$.

\section{Action and field equations}
Taking the viewpoint of effective field theory,
we consider the action as a derivative expansion,
keeping all terms consistent with diffeomorphism
symmetry. The most general Lagrangian scalar density
involving the metric $g_{ab}$ and preferred frame
unit vector $u^a$ with two
or fewer derivatives is $\tilde{\cal
L}=\sqrt{-g}{\cal L}$, with
\beq {\cal L} =  -R - K^{ab}{}_{mn}(\nabla_a u^m)(\nabla_b
u^n)-\l(g_{ab}u^au^b-1) \label{action} \eeq
where
\beq K^{ab}{}_{mn} = c_1 g^{ab}g_{mn}+c_2 \d^a_m\d^b_n +c_3
\d^a_n\d^b_m+c_4 u^au^bg_{mn}. 
\label{K}
\eeq
Stationarity under variation of the Lagrange multiplier $\lambda$
constrains the preferred frame vector $u^a$ to be unit
timelike.
We have omitted in the lagrangian terms that would
vanish when the unit constraint is satisfied.
The unit timelike vector $u^a$ is present everywhere in
spacetime in every field configuration and specifies a
locally preferred rest frame. It can be thought of as the
four-velocity of a ubiquitous fluid and hence is naturally
called the {\it aether} field. We sometimes use the term
{\it aether theory} to refer to the theory described by the
Lagrangian (\ref{action}).

A general class of Lagrangians for vector-tensor theories with an 
unconstrained vector was parametrized by Will and 
Nordvedt~\cite{willnord,willbook}. Their term
$\omega u^a u_a R$ does not appear here because of the unit constraint,
and our $c_4$ term does not appear there.
The relation between the parameters in Ref.~\cite{willbook}
(neglecting $\omega$) and ours (neglecting $c_4$) is
\beq
\begin{array}{ll}
\tau = -(c_1+c_2+c_3) \qquad \qquad& c_1 = 2\epsilon-\tau\\
\eta = -c_2 & c_2 = -\eta\\
\epsilon = -(c_2+c_3)/2 & c_3 = \eta-2\epsilon
\end{array}
\eeq

In order to agree with observations the dimensionless coefficients
$c_{1,2,3,4}$ must presumably be fairly small compared to
unity. Special cases of this action are identified in Table \ref{cases}.
\begin{table*}
\caption{\label{cases}Special cases of the action (\ref{action}).}
\begin{ruledtabular}
\begin{tabular}{lll}
Special case&parameter values in (\ref{K})&parameter values in~\cite{willbook}\\
\hline
General Relativity&$c_1=c_2=c_3=c_4=0$&  $\tau=\eta=\epsilon=0$ \\
 equivalent to GR by field redefinition~\cite{Barbero}&$c_1+c_4=0$,\quad $c_2+c_3=0$ &$\tau=\epsilon=0$\\
Will-Nordtvedt~\cite{willnord, willbook,mattjacob}&
$c_2=c_3=c_4=0$&$\eta=\epsilon=0$\\
Hellings-Nordvedt~\cite{hellnord}& $c_1+c_2+c_3=0$,\quad $c_4=0$& $\tau=0$\\
Einstein-Maxwell-like~\cite{KosSam,jacobmatt}&$c_1+c_3=0$,\quad $c_2=c_4=0$& $\tau=\eta=0$
\end{tabular}
\end{ruledtabular}
\end{table*}
Note that the ``Einstein-Maxwell-like" case is a
sub-case of Hellings-Nordtvedt. This case has
an extra gauge symmetry and was
disfavored in \cite{jacobmatt} on account of the
gradient singularities that generally develop in the
vector field. Another notable
sub-case of Hellings-Nordtvedt is
$c_1=c_4=0$,\quad $c_2+c_3=0$, which is
also one of the theories equivalent to GR via a field redefinition.

The metric equation with no matter source
(other that the aether field) can be written in the form
\beq
G_{ab} = T_{ab}\label{AEE}
\eeq
where $G_{ab}$ is the Einstein tensor and $T_{ab}$ is the ``aether
stress" tensor obtained from varying the aether part of the action
(\ref{action}) with respect to the metric. For the metric
variation we use the inverse metric, but we also have a choice
whether to consider the independent aether field to be a covariant
or a contravariant vector. We choose contravariant $u^a$ to
simplify the stress tensor a bit, since the action in the
contravariant form (\ref{action}) has no metric dependence
associated with the $c_2$ and $c_3$ terms in $K^{ab}{}_{mn}$. The
field equations are thus obtained by requiring that the action
(\ref{action}) be stationary with respect to variations of
$g^{ab}$, $u^a$, and $\l$.

The $\lambda$ variation imposes the unit constraint,
\beq g_{ab} u^a u^b = 1.  \label{lambdafield} \eeq
The $u^a$ variation gives
\beq \nab_a J^{a}{}_m-c_4 \dot{u}_a \nab_m u^a = \l u_m,
\label{ueqn}\eeq
where to compactify the notation we have defined
\beq J^{a}{}_{m} = K^{ab}{}_{mn} \nabla_b u^n \eeq
and
\beq \dot{u}^m = u^a\nabla_a u^m. \eeq
Solving for $\lambda$ using (\ref{lambdafield}) we find
\beq \l=u^m \nab_a J^{a}{}_{m} - c_4 \dot{u}^2. \label{lambda} \eeq
The $g^{ab}$ variation yields the aether stress tensor
\begin{eqnarray} 
T_{ab}&=&
\nab_m(J_{(a}{}^m u_{b)} - J^m{}_{(a} u_{b)} - J_{(ab)}u^m) \nonumber\\
&&+ c_1\, \left[(\nab_m u_a)(\nab^m u_b) -(\nab_a u_m)(\nab_b u^m) \right]\nonumber\\ 
&&+ c_4\, \dot{u}_a\dot{u}_b\nonumber\\
&&+\left[u_n(\nab_m J^{mn})-c_4\dot{u}^2\right]u_a u_b \nonumber\\
&&-\frac{1}{2} g_{ab}{\cal L}_u.
\label{aetherT}
\end{eqnarray}
In the above expression (\ref{lambdafield}) has been used
to eliminate the term that arises
from varying $\sqrt{-g}$ in the constraint term in
(\ref{action}),
and in the fourth line $\lambda$ has been eliminated using
(\ref{lambda}). The first line
contains all of the terms arising from varying the
metric dependence of the connection. Note that it contains
terms of second order in derivatives. In the last line
the notation ${\cal L}_u$ refers to all of ${\cal L}$ in (\ref{action})
except the Ricci scalar term.

\section{Spherically symmetric static solutions}
\label{Symm}

Our objective in this paper is to consider the weak field
limit of spherically
symmetric static solutions to the aether-metric field equations.
In spherical symmetry the $c_4$ term in the action
can be absorbed by the change of
coefficients
\bea
c_1&\rightarrow& c_1+c_4\nonumber\\
c_3&\rightarrow& c_3-c_4.
\label{absorbc4}
\eea
To see why, note that
any spherically symmetric vector field is
hypersurface orthogonal, hence the twist
\beq \omega_a = \epsilon_{abcd}u^{b} \nabla^{c} u^{d}\eeq
of the aether vanishes.  The identity~\cite{Barbero}
\beq  \dot{u}^2= -\omega_a \omega^a + \nabla_a u_b \nabla^a u^b -
\nabla_a u_b \nabla^b u^a, \label{bidentity}\eeq
valid
for $u$ satisfying $u^2=1$, can be used to trade
the $\dot{u}^2$ term in the action (\ref{action}) for an $\omega^2$ term
together with the substitution (\ref{absorbc4}).
Since the twist occurs quadratically and vanishes
in spherical symmetry, that term will not contribute to the field
equations, hence the $c_4 \dot{u}^2$ term simply modifies
the coefficients as indicated in (\ref{absorbc4}). Thus 
we henceforth set $c_4=0$ without loss of
generality, as 
it can be reintroduced at the end via the
replacements (\ref{absorbc4}). 
(Although substitution
of the identity (\ref{bidentity}) 
will not change the content of
the field equations, it will change the
value of the lagrange multiplier $\l$ for a given solution.)

We have analyzed the
asymptotic limit of such solutions and found
at first PPN order the two ERS parameters
are exactly the same as in pure GR
as long as $c_1+c_2+c_3 \ne 0$. 
The special case $c_1+c_2+c_3=0$ has no single characterization.
We now describe how these results are obtained.

\subsection{Field equations}
A common choice for a weak field analysis is
isotropic coordinates $(t,r,\theta,\phi)$, which we adopt
here.
In these coordinates the line element is
\beq ds^2 = N(r) dt^2 - B(r)(dr^2+r^2d\Omega^2). \eeq
(Note that $r$ is not
the usual Schwarzschild radial coordinate.)
The aether field takes the form
\beq u^t(r)\frac{\partial}{\partial t} + u^r(r)
\frac{\partial}{\partial r} = a(r)\frac{\partial}{\partial t} +
b(r)\frac{\partial}{\partial r} \label{uform}\eeq
and the unit constraint becomes \beq N(r)a(r)^2-B(r)b(r)^2 = 1.
\label{constr} \eeq
The aether field equation (\ref{ueqn})
has just $t$ and $r$ components,
and the elimination of $\l$ reduces this pair to
one independent equation.

Solving the field equations is obviously an enormous task given
the form (\ref{connection}--\ref{aetherT}) 
of the stress tensor so we used the
symbolic math program MAPLE and the Riemann tensor
package~\cite{riemannp}. With this package one can easily express
the field equations in terms of the functions
$N(r),B(r),a(r),b(r)$ and their derivatives. The end result is a
set of coupled ordinary differential equations coming from both
the Einstein equation and the aether field equation. Given that
there are only three free functions left after applying the
constraint (\ref{constr}), just three independent ODE's are needed.
We used the aether field equation (\ref{ueqn}) and the $tt$
and $rr$ components of the metric equation. 
The equations are sufficiently complicated that
it does not seem illuminating to display them here.

\subsection{Asymptotic weak field limit} \label{WeakLim}

Far from the source, the metric should approach flat Minkowski space.
In order to examine what happens as $r$ approaches infinity
we introduce the change of variables
\beq x = \frac{1}{r}. \eeq
Around $x = 0$ the functions $N(x), B(x),
b(x)$ will have power series behavior in the form of
\begin{eqnarray}
 N(x)&=&1+N_1x+N_2 x^2+N_3 x^3+N_4 x^4\\B(x)&=&1+B_1x+B_2x^2+B_3 x^3+B_4 x^4\\
 b(x)&=&b_0+b_1x+b_2x^2+b_3 x^3+ b_4 x^4.
\end{eqnarray}
At this stage it is convenient to use the constraint equation
(\ref{constr}) to eliminate $a(r)$ in favor of the radial
component $b(r)$. It turns out that asymptotic flatness and
spherical symmetry generally require the aether to have no radial
component at infinity ($a_0 = 1$, $b_0 = 0$) except in
the Einstein-Maxwell-like case  where
the action takes a special form with an additional 
symmetry. The first
order coefficient $N_1$ determines the Newtonian gravitational
potential, so what we are really interested in are the
post-Newtonian corrections to this associated with the $B_1$ and
$N_2$ coefficients. The higher order coefficients are post-post
Newtonian (and beyond). Substituting the above forms of the
functions into the equations of motion and performing a series
expansion in Maple around the point $x = 0$ ultimately gives a set
of algebraic equations that can be solved to produce the local
power series solutions for the fields.

\subsection{Series solutions}
\label{Solutions}

To illustrate our methods we first discuss the local
power series solutions to pure Einstein gravity in isotropic
coordinates. In this case, the $c_{i}$ parameters are all set to
zero and we are left to consider the two coupled ODE's for the
functions $N(x)$, $B(x)$ given by the vanishing
$tt$ and $rr$ components of
the Einstein tensor. Using the procedure described in section
\ref{WeakLim} we find
\begin{eqnarray}
G_{tt} &=& \left[\frac{3}{4}B_1^2-2 B_2\right]x^4
+\left[N_1\left(\frac{3}{4}B_1^2-2 B_2\right)\right. \nonumber\\&&
\left. -\frac{9}{4}B_1^3+7 B_1B_2-6B_3\right] x^5+ \cdots\\
G_{rr} &=& -(N_1+B_1)x^3
+\left[N_1^2 +\frac{1}{2}N_1B_1\right. \nonumber\\&& 
\left. +\frac{5}{4}B_1^2 -2N_2-2B_2\right]
x^4 + \cdots
\end{eqnarray}
Thus, at third order in $x$, the ${rr}$ equation implies
\beq B_1 = -N_1. \eeq
We can then substitute this result into the fourth order equations
and solve simultaneously to determine that
\begin{eqnarray}
N_2 &=& \frac{1}{2} N_1^2\\
B_2 &=& \frac{3}{8} N_1^2.
\end{eqnarray}
For the additional coefficients in $B(x)$ and $N(x)$ we continue
the process of examining higher order equations, substituting in
lower order results, and solving simultaneously. These solutions
can of course be verified by simply expanding the
commonly known solution for the functions in isotropic
coordinates in a power series.

Now we return to the case of interest and tune the $c_{i}$
parameters back up to non-zero values. At lowest (second) order in
$x$ the aether field equation tells us that
\beq -2 (c_1+c_2+c_3) b_0 (b_0^2+1) = 0, \label{eqn1} \eeq
which says that $b_0 = 0$ provided that $c_1+c_2+c_3 \ne 0$. This
combination of parameters also appears in the aether field
equation at third order in $x$,
 \beq
-2(c_1+c_2+c_3) b_1 = 0. \label{eqn2}
 \eeq
From equations (\ref{eqn1}) and (\ref{eqn2}) it is clear that we
have two completely different cases depending on whether
or not
$c_1+c_2+c_3$ vanishes.  These cases must be analyzed separately.

\subsection{Generic case: $c_1+c_2+c_3 \ne 0$}
\label{nonzero}

For this generic case (\ref{eqn1}) shows that
asymptotic flatness of the
metric implies that $b_0=0$, i.e. the aether has no radial
velocity at infinity.
Together with the constraint this implies that 
$a_1 = -N_1/2.$ In addition, (\ref{eqn2}) tells us 
that $b_1 = 0$.

Now let us consider the metric equations.
The $rr$ equation tells us that
\beq
B_1 = -N_1,
\label{gamma}
\eeq
a
result identical to pure GR. We have now determined all of the
zeroth and first order coefficients in terms of $N_1$, but
to examine the higher order ones we must consider the higher order
terms in the expansions of the field equations. At fourth order the
$u$ field equation is identically zero after substituting $b_1 =
0$ and $B_1=-N_1$. Now all that remains at this order is to
determine $B_2$ and $N_2$ using the two Einstein equations at fourth
order in $x$. These have the form
\begin{eqnarray}
\frac{5}{8} N_1^2 c_1-c_1 N_2-2
B_2+\frac{3}{4} N_1^2 &=& 0\\
\frac{7}{4}
N_1^2-2 N_2-2 B_2+\frac{1}{8} N_1^2 c_1 &=& 0.
\end{eqnarray}
Solving these two equations simultaneously yields the final result
\begin{eqnarray}
N_2 &=& \frac{1}{2} N_1^2\label{beta}\label{N2}\\ 
B_2&=&\frac{3}{8} N_1^2+\frac{1}{16} N_1^2 c_1
\end{eqnarray}

To determine further coefficients of the power series expansion we
move on to consider the field equations at fifth and sixth order in
$x$. At fifth order in the $u$ field equation we recover the result
\beq
(c_1+c_2+c_3)(b_2 N_1-b_3) = 0
\eeq
indicating that $b_2$ is a new free parameter in addition to $N_1$,
and $b_3 = N_1 b_2$. The remaining metric equations at fifth order
are quite complicated so we simply quote the final results
\begin{eqnarray}
B_3 &=& -\frac{1}{16}N_1^3-\frac{5}{96}N_1^3 c_1\\
N_3 &=&\frac{3}{16}N_1^3-\frac{1}{96}N_1^3 c_1
\end{eqnarray}
We also examined the sixth order equations to find $N_4$ and
$B_4$, but we will not give the results due to their complexity.
However, we note that these coefficients depend on both $N_1$ and
$b_2$.

As a final note, we also expanded the equation for lambda in
(\ref{lambda}) and used all of the above results for the expansion
coefficients to determine at what order lambda contributes. This
yields
\beq
\lambda = \frac{1}{2}N_1^2 c_1 x^4 + \cdots
 \eeq
(As mentioned 
at the beginning of this section, 
the $c_4$ dependence of $\l$ cannot be obtained via
the substitutions (\ref{absorbc4}).) 

\subsection{Special case: $c_1+c_2+c_3 = 0$}
\label{zero}

This special case corresponds to the Hellings-Nordtvedt theory~\cite{hellnord}
with a unit constraint on the vector field.
Setting $c_3 = -c_2-c_1$ from the beginning and repeating the
procedure we find that the second order $tt$ and $rr$ metric
equations imply
\beq
c_2 b_0^2 = 0.
\eeq
This special case thus further subdivides into the cases
$c_2\ne0$ and $c_2=0$.

\subsubsection{$c_2\ne0$}
If $c_2$ is non-zero we again find that $u^a$ has no radial
component at infinity. There is no single characterization of this
case. An exceptional  sub-case occurs if $c_1=0$, which falls into
the class~\cite{Barbero} that is equivalent to GR via a field
redefinition (and the lagrangian is just $R+c_2R_{ab}u^au^b$). 
In  this class the aether field is completely
unconstrained. If $c_1\ne0$ we again find $B_1=-N_1$ as in (\ref{gamma}),
while unlike (\ref{N2}) we find 
\beq
N_2 = \frac{1}{8} N_1^2 (c_1+6c_2+4)/(c_2+1).
\label{unlike} 
\eeq

\subsubsection{$c_2 = 0$}

The case $c_2 = 0$ yields the
Einstein-Maxwell-like (plus $c_4$) sector of the
theory, which was previously
analyzed non-perturbatively in \cite{jacobmatt}.
Working through the procedure for finding the local power
series solutions we find the Reissner-Nordstrom solution (provided
$\lambda = 0$) with $b_0$, $b_1$, and $N_1$ as free parameters.
The freedom appearing here in $b_0$ and $b_1$ is a result of the
limited gauge symmetry
\beq
u_a\rightarrow u_a + \nabla_a f
\eeq
preserving the unit constraint, as
discussed in \cite{jacobmatt}. Specifically,
$b_1$ is associated with an ``aether charge" while $b_0$
corresponds to a scaling freedom. This is similar to the usual
Reissner-Nordstrom case where the general solution for the
co-vector potential $A_t$ is
\beq
A_t = \frac{Q+Dr}{r},
\eeq
where the $D$ constant is usually set to zero so that the field
will be 0 at infinity.

The solutions with $\l\ne0$ have the aether aligned with the
Killing vector, i.e.  $b(r)\equiv0$. While there always exist such
solutions in this special case, they are not asymptotically flat
except in the even more special case $c_1=-c_3=2$, $c_2=0$. (In
that case there is a full functional freedom in the solution,
which corresponds in the charged dust interpretation of
\cite{jacobmatt} to the case of extremally charged
dust.) Thus the exterior solution for a star must have
$\l=0$. On the other hand at the origin we must have 
$\l\ne0$ to avoid a $1/r$ singularity in the $u$-field.
It does not appear possible to match these, so it may
be that there are no static spherically symmetric 
solutions that are regular at the origin. Since the
Einstein-Maxwell case was already deemed unphysical~\cite{jacobmatt}
due to the generic appearance of aether shocks, we shall
not belabor this point here.

\subsection{Eddington-Robertson-Schiff parameters}

In the usual analysis of the post-Newtonian corrections to the
gravitational field of a static spherical body
the Schwarzschild line element is rewritten in terms of
isotropic coordinates and those metric coefficients are then
expanded to post-Newtonian accuracy. This takes the following form
for a general gravitational theory~\cite{willbook}:
\beq ds^2 = \Bigl(1-\frac{2M}{r}+ 2 \beta\frac{M^2}{r^2}\Bigr)dt^2 -
\Bigl(1-2\gamma\frac{M}{r}\Bigr)[dr^2 + r^2d\Omega^2]
\eeq
where $M$ is the gravitating mass of the body in geometric units
and $\gamma$ and $\beta$ are the
Eddington-Robertson-Schiff (ERS) parameters of the theory.
The parameter $\gamma$ measures
the amount of space curvature produced by a
unit rest mass and $\beta$ describes the amount of non-linearity
in the superposition law. 

In the generic case $c_1+c_2+c_3\ne0$, we read off
from equations (\ref{gamma}) and (\ref{beta})
of section \ref{nonzero} that
 \begin{eqnarray}
 \gamma &=& 1\\\beta &=& 1,
\label{genericERS}
 \end{eqnarray}
in exact agreement with pure GR. 
The special case  $c_1+c_2+c_3=0$
has no single characterization.
If $c_1=0$ it is equivalent to GR via 
a field redefinition, and the aether field is arbitrary.
If $c_2=0$ it is the Einstein-Maxwell-like sector,
and the exterior is described by the Reissner-Nordstrom 
solution.  Hence $\g=1$, but the value of $\b$ depends upon
the aether charge which is not determined by our method.
If neither $c_1$ nor $c_2$ vanishes then, using (\ref{unlike})
we find 
\begin{eqnarray}
 \gamma &=& 1\\\beta &=& \frac{1}{4}\frac{c_1+6c_2+4}{c_2+1},
 \end{eqnarray}
where (\ref{unlike}) was used to obtain $\b$. This special
case corresponds to $\t=\o=0\ne\eta$ which,
as shown by Will~\cite{willbook}, is dynamically 
overdetermined in the linearized, unconstrained vector-tensor theory. 

\section{Discussion}
\label{Discussion}

There are two important implications of this analysis. First,
there appear to be only two free parameters in the local solution
around infinity for the generic case  $c_1+c_2+c_3 \ne 0$, namely
$N_1$ and $b_2$. It is possible that analyzing the global behavior
of the field equations may eliminate one of these or demonstrate
the existence of even more parameters. Based on an analogy from
pure GR, the metric parameter $N_1$ is determined by the mass of
the presumed static, central object generating the field. The
aether parameter $b_2$ cannot be associated with a ``charge" as in
the special case of Einstein-Maxwell due to the $1/{r^2}$ fall
off.

The second implication is that in the generic case
the aether model is quite close
observationally to pure GR since the ERS parameters
match.  More precisely, the coefficients of
the metric expansions are identical up to $B_2$, which
differs by a term of relative size
$c_1$ (or $c_1+c_4$).
Other alternative
theories of gravity with the same ERS parameters
are the general vector-tensor theory without
the unit constraint\footnote{This agreement
is obtained upon setting the term $\omega K^aK_a R$ to zero
in \cite{willbook}, on the grounds that all it does is
renormalize $G$ given the unit constraint on $K^a$.}, 
and the bimetric theories with prior geometry
of Rosen and of Rastall~\cite{willbook}.
The fact that
the ERS parameters are the same suggests that there
may be a closer relation than might be expected
between the PPN parameters
of unconstrained and constrained vector tensor theories.
It also is interesting to note the differences between
this model and the Brans-Dicke scalar-tensor theory. The
Brans-Dicke parameters are
\begin{eqnarray}
\gamma &=& \frac{1+\omega}{2+\omega}\\
 \beta &=& 1
 \end{eqnarray}
where $\omega$ is the Dicke coupling constant, which must be greater
than 500 in order to agree with observation.

In order to have a comprehensive check on the theory in the solar
system we need to consider the full post-Newtonian approximation
scheme. This allows for preferred frame effects due to the
motion of the solar system with respect to the asymptotic preferred
frame and is described by ten parameters (two of which are $\gamma$ and
$\beta$). To determine these parameters one perturbatively
integrates the field equations with a fluid source, imposing the condition of
regularity at the origin. It seems likely that this would  fix the value of
$b_2$ in the generic case.
Further tests of preferred frame effects will be
found in gravitational wave phenomena (briefly mentioned in
\cite{mattjacob}) such as the orbital decay of binary pulsars,
and in strong field settings such as black holes.

\section*{Acknowledgments}
This work was supported in part by the NSF under grants
PHY-9800967 and PHY-0300710 at the University of Maryland.


\begin{thebibliography}{99}


\bibitem{willnord}C.M.~Will and K.~Nordvedt, Jr., ``Conservation laws
and preferred frames in relativistic gravity. I. Preferred frame theories and
an extended PPN formalism, Astrophys.\  J.\ {\bf 177}, 757 (1972).

\bibitem{nordwill}K.~Nordvedt, Jr. and C.M.~Will, ``Conservation laws
and preferred frames in relativistic gravity. II. Experimental evidence
to rule out preferred frame theories of gravity",
Astrophys.\  J.\ {\bf 177}, 775 (1972).

\bibitem{hellnord}R.W.~Hellings and K.~Nordvedt, Jr., ``Vector-metric
theory of gravity", Phys.\ Rev.\  {\bf D7}, 3593 (1973).

\bibitem{willbook}C.M. Will, {\it Theory and Experiment in Gravitational
Physics}, (Cambridge Univ. Press, Cambridge, 1993).

\bibitem{moffatv}M.A.~Clayton and J.W.~Moffat,
``Dynamical mechanism for varying light velocity as a solution to
cosmological problems,''
Phys.\ Lett.\  {\bf B460}, 263 (1999).

\bibitem{moffats}M.A.~Clayton and J.W.~Moffat,
``Scalar-tensor gravity theory for dynamical light velocity,''
Phys.\ Lett.\  {\bf B477}, 269 (2000).

\bibitem{Bassett:2000wj}
B.A.~Bassett, S.~Liberati, C.~Molina-Paris and M.~Visser,
``Geometrodynamics of variable speed of light cosmologies,''
Phys.\ Rev.\ D {\bf 62}, 103518 (2000)
[arXiv:astro-ph/0001441].

\bibitem{KosSam}V.A.~Kosteleck\'y and S.~Samuel,
``Gravitational phenomenology in higher dimensional theories and
strings,'' Phys.\ Rev.\  {\bf D40}, 1886 (1989).

\bibitem{jacobmatt}T.~Jacobson and D.~Mattingly, ``Gravity with a dynamical
preferred frame,'' Phys.\ Rev.\  {\bf D64}, 024028 (2001).

\bibitem{mattjacob}D.~Mattingly and T.~Jacobson, ``Relativistic
gravity with dynamical preferred frame,'' in {\it CPT and Lorentz
Symmetry II}, ed. V.A. Kostelecky (World Scientific, Singapore,
2002) [arXiv:gr-qc/0112012].

\bibitem{Gasperini}See, for example, M.~Gasperini,
``Singularity prevention and broken Lorentz symmetry",
Class.\ Quantum\ Grav.\ {\bf 4}, 485 (1987);
``Repulsive gravity in the very early Universe",
Gen. \ Rel.\ Grav.\ {\bf 30}, 1703 (1998); and
references therein.

\bibitem{Barbero}
J.~F.~Barbero~G. and E.~J.~Villase\~{n}or,
``Lorentz Violations and Euclidean Signature Metrics,''
Phys.\ Rev.\  {\bf D68}, 087501 (2003).

\bibitem{riemannp}R.~Portugal and S.L.~Sautu, ``Applications of
Maple to General Relativity,'' Computer Physics Communications
{\bf 105} (1997). Also see
http://www.cbpf.br/$\sim$portugal/Riemann.html.


\end{thebibliography}
\end{document}